\documentclass[preprintnumbers,amsmath,amssymbm,prd]{revtex4}
\usepackage{epsfig}
\usepackage{graphicx}
\usepackage{xcolor}

\begin{document}
\title{The third law of black hole thermodynamics: Quantum versus classical considerations}
\author{Shahar Hod$^1$ and Tsvi Piran$^2$}
\affiliation{$^1$The Ruppin Academic Center, Emeq Hefer 40250, Israel}
\affiliation{$^2$Racah Institute of Physics, The Hebrew University, Jerusalem 91904, Israel}
\date{\today}

\begin{abstract}

Kehle and Unger have recently shown that, under suitable conditions, a massless charged classical scalar field can produce an extremal black hole in finite time. This result appears to violate the third law of thermodynamics (TLT), since extremal black holes have zero temperature. However, because the TLT is intrinsically a quantum principle, a self-consistent analysis of this problem must account for quantum effects. 
In particular, we point out that the electric field of the system may spontaneously polarize the vacuum, 
thereby creating massless charged pairs that will discharge the black hole. 
This mechanism guarantees the {\it quantum} validity of the TLT in the regime $|eQ|\geq\hbar/2$ 
(here ${Q,e}$ denote, respectively, the electric charges of the black hole and the scalar field). 
This leaves open the question of what happens for lower values of $|eQ|/\hbar$. 
Recently, Schneider derived a classical bound showing that extremal black holes cannot be formed \`a la Kehle-Unger 
in the small-charge regime $|eQ|\leq\hbar/3$. His analysis leaves open the possibility that a stronger bound may exist. 
We conjecture here that this bound can indeed be strengthened such that classical considerations would prevent extremal (zero-temperature) charged Reissner-Nordstr\"om black holes from forming dynamically in the regime $|eQ|<\hbar/2$, where quantum polarization effects are too weak to preserve the validity of the TLT.
\end{abstract}
\bigskip
\maketitle

\section{Introduction}

Is it possible to reach absolute zero temperature in a finite time? 
This fundamental question has attracted the attention of physicists for over a century. 
An explicit answer to this intriguing question is provided by the third law of thermodynamics (TLT), 
which asserts that it is impossible for any physically self-consistent procedure to reach absolute zero 
temperature $T=0$ in a finite number of steps \cite{TL}, or, equivalently, in a finite advanced time.

In the 1970s, Bekenstein provided compelling arguments that, within the framework of a 
semiclassical theory (with $\hbar\neq0$), black holes possess a finite entropy, $S_{\text{BH}}\propto \text{Area}/\hbar$ \cite{Bk73,Bk74} (we use gravitational units in which $G=c=1$). Bardeen, Carter, and Hawking subsequently showed that black holes obey four mechanical laws analogous to the laws of thermodynamics \cite{BCH}. 
However, it was only after Hawking demonstrated \cite{H75} that black holes evaporate by emitting thermal radiation at a temperature proportional to their surface gravity that the mechanical laws of black-hole mechanics \cite{BCH} became widely accepted as genuine thermodynamic laws, with the proportionality factor in the entropy-area 
relation equal to $1/4$.

A rigorous mathematical proof of the classical black-hole version of the second law -- 
the statement that, under physically plausible conditions, the horizon area of a black hole cannot decrease -- 
had already appeared in the physics literature \cite{Ch70,CR71,Ha71} even before the realization that black holes possess thermodynamic properties. 
Bekenstein \cite{Bk81,Bk83} and Unruh and Wald \cite{UW82,UW83} subsequently revealed that the generalized second law (GSL) of thermodynamics \cite{Bk73,Bk74} is intrinsically a quantum law.
In particular, it was explicitly demonstrated in \cite{Bk81,Bk83,UW82,UW83} that quantum considerations must be taken into account for the GSL to hold in the presence of black holes.

In analogy with the classical arguments \cite{Ch70,CR71,Ha71} leading to 
the generalized second law \cite{Bk73,Bk74}, 
Israel \cite{Is86} presented a classical proof of the third law of black-hole thermodynamics. 
Recently, Kehle and Unger \cite{KU} identified a loophole in Israel's celebrated proof. They showed that, within the classical framework of the coupled Einstein-Maxwell-charged-massless-scalar field theory, there exist solutions describing the dynamical formation of exactly extremal (zero-temperature) Reissner-Nordstr\"om black holes in finite advanced time. 
It was therefore argued in \cite{KU} that this collapse model provides a definitive violation of the third law of black-hole thermodynamics.
Motivated by the results of Kehle and Unger \cite{KU}, we raise here the following question:
Is there a physical mechanism that can prevent the apparent Kehle-Unger violation of the TLT in charged black-hole spacetimes?

\section{Quantum effects and the TLT}

The TLT, like the GSL, is intrinsically a {\it quantum} law \cite{Hodbo,HodQ2,Gru,Pes,HodQ3}. 
One therefore expects {\it quantum}-mechanical considerations to play an essential role in ensuring its validity. 
One important mechanism that may protect 
the TLT in collapse scenarios involving self-gravitating charged fields is the Schwinger-type pair-production mechanism \cite{Schw1,Schw2,Schw3,Schw4}, a purely quantum effect that may 
constrain the electric field strength of the charged configuration.
In particular, in the composed Einstein-Maxwell-charged-massive-scalar field theory, vacuum polarization effects 
impose the quantum upper bound \cite{Schw1,Schw2,Schw3,Schw4}
\begin{equation}\label{Eq1}
E<E_{\text{c}}\equiv{{m^2}\over{e\hbar}}
\end{equation}
on the electric field strength of the charged configuration, where $m$ is the proper mass of the charged field. 

However, for a charged configuration with an effective radius $R$, the Schwinger upper bound (\ref{Eq1}) is valid only in the small Compton wavelength regime $\hbar/m\ll R$, where the electric field can be regarded as approximately constant. 
In their collapse model, Kehle and Unger \cite{KU} considered the opposite limit $mR\ll\hbar$, corresponding to massless (or very light) scalar fields. 
In this regime, the Schwinger bound in its simplest form (\ref{Eq1}), 
which is based on the assumption of a constant electric field strength, is no longer applicable.

In the regime relevant to \cite{KU}, $mR\ll\hbar$, Flambaum and Berengut \cite{Flam} have shown 
that if the dimensionless charge parameter $eQ/\hbar$ of the charged configuration exceeds a critical value, 
the electric field can spontaneously polarize the vacuum and produce pairs of oppositely charged massless particles. 
This provides a quantum mechanism that, in the $mR\ll\hbar$ regime, 
bounds from above the charge parameter $eQ/\hbar$ of the system 
(we shall assume $Q,e>0$ without loss of generality). 

The critical value of the dimensionless charge parameter $eQ/\hbar$ for the production 
of massless charged particles can be estimated using the following heuristic argument: 
Suppose that a central source with electric charge $Q>0$ polarizes the vacuum, creating a pair of oppositely charged massless particles. The positively charged particle escapes to infinity with energy
\begin{equation}\label{Eq2}
E^{+}\geq0\
\end{equation}
while the negatively charged particle remains bounded to the central positively charged source. 
At a distance $\sim r$ from the source, the Coulomb potential energy 
is $E^{-}_{\text{elec}}(r)\sim-|eQ|/r$. 
In addition, localizing a massless particle to a region of size $\sim r$ gives, 
by the uncertainty principle, a quantum-mechanical 
energy $E^{-}_{\text{qm}}(r)\gtrsim\hbar/(2r)$. Thus, the energy of the negatively charged bounded particle is characterized by the relation
\begin{equation}\label{Eq3}
E^{-}(r)\gtrsim-{{|eQ|}\over{r}}+{{\hbar}\over{2r}}\ .
\end{equation}

Pair production becomes energetically possible when the electric field can supply an energy comparable to the 
quantum-mechanical energy scale. In particular, energy conservation requires $E^{+}+E^{-}=0$, which implies
\begin{equation}\label{Eq4}
-{{|eQ|}\over{r}}+{{\hbar}\over{2r}}\lesssim0\ .
\end{equation}
The dependence on the distance scale $r$ then cancels, yielding the threshold condition
\begin{equation}\label{Eq5}
\Big({{eQ}\over{\hbar}}\Big)_{\text{crit}}\simeq{1\over2}\  .
\end{equation}
Once this condition is satisfied, the electric field becomes sufficiently strong to pull oppositely charged massless particles out of the vacuum. The two members of the pair are accelerated in opposite directions by the background electric field, thereby providing a mechanism for discharging the source. 
(Note that this heuristic argument also applies to a charged field with proper mass $m$ whose Compton wavelength $\hbar/m$ is much larger than the characteristic size of the charged source).

By solving the Klein-Gordon equation for a charged scalar field with proper mass $m$ and electric charge $e$ in the background of a charged configuration with an effective radius $R\ll \hbar/m$ and electric charge $Q$, 
one finds that the ground state of the field reaches the lower continuum (corresponding to the energy $E=-m$) when \cite{Flam,Noter0}
\begin{equation}\label{Eq6}
\Big({{eQ}\over{\hbar}}\Big)_{\text{crit}}={1\over2}+O[\ln^{-2}(\hbar/mR)]\ \ \ \ \text{for}\ \ \ \ mR\ll\hbar\  .
\end{equation}
When the dimensionless charge parameter $eQ/\hbar$ reaches the critical value (\ref{Eq6}), 
the binding energy of the system is $2m$, rendering it unstable against spontaneous vacuum polarization. The electric field can then spontaneously polarize the vacuum and create a pair of oppositely charged particles: the negatively charged particle is captured by the positively charged configuration (thereby decreasing its effective electric charge), while the positively charged particle escapes to infinity. 
Note that, at the threshold (\ref{Eq6}), the captured negatively charged particle carries vanishing energy, so that the capture process reduces the charge of the configuration while leaving its mass essentially unchanged, thereby driving the system away from extremality.

Remarkably, it was shown in \cite{Kt} that, for near-extremal Reissner-Nordstr\"om black holes, 
spontaneous pair production occurs in the regime $(eQ/\hbar)^2>1/4+(mQ/\hbar)^2$, which, for charged massless fields, agrees with the flat-space relation (\ref{Eq6}), yielding the critical value 
$(eQ/\hbar)_{\text{crit}}=1/2$ for $m=0$.

One therefore concludes that, in the regime $eQ/\hbar\geq (eQ/\hbar)_{\text{crit}}$ of the coupled 
Einstein-Maxwell-massless-scalar field theory, 
quantum effects polarize the vacuum, thereby reducing the effective charge of the self-gravitating 
configuration before it can form an extremal black hole:
\begin{equation}\label{Eq7}
{{eQ}\over{\hbar}}\geq{1\over2}\ \ \ \Longrightarrow\ \ \ \text{Quantum effects prevent
the dynamical formation of extremal black holes.}
\end{equation}
Thus, for $eQ/\hbar \geq 1/2$, even if extremal charged black holes can be formed classically, 
this does not imply a violation of the TLT. 

\section{A conjectured bound on the formation of extremal black holes in the classical 
Einstein-Maxwell-massless-scalar system}

While quantum discharge is effective in the regime $eQ/\hbar \geq 1/2$, one may 
wonder whether the TLT is violated for $eQ/\hbar<1/2$. Remarkably, classical effects seem to save the day. 
Schneider recently published a mathematical theorem \cite{Schn} showing that, within the framework of the classical Einstein-Maxwell-massless-scalar field theory, the dynamical formation of 
extremal black holes \`a la Kehle-Unger can occur only above a certain critical dimensionless electric charge:
\begin{equation}\label{Eq8}
{{eQ}\over{\hbar}}>\Big({{eQ}\over{\hbar}}\Big)_{\text{min}}\ .
\end{equation}
The exact value of $({{eQ}/{\hbar}})_{\text{min}}$ is currently unknown, but it has been shown to lie 
within the narrow interval \cite{Schn} 
\begin{equation}\label{Eq9}
\Big({{eQ}\over{\hbar}}\Big)_{\text{min}}\in\Bigg[{1\over3},\sqrt{{{3+\sqrt{33}}}\over{3}}\Bigg]\ .
\end{equation}
The lower limit of the interval (\ref{Eq9}) containing the exact value of $(eQ/\hbar)_{\text{min}}$ is not known to be sharp, whereas the upper limit is sharp in the sense that Schneider provides an explicit example in which an extremal black hole is dynamically produced for $eQ/\hbar>\sqrt{(3+\sqrt{33})/3}$. 
Taking into account the Schneider relation (\ref{Eq9}) and the quantum bound (\ref{Eq7}), 
one deduces that any possible violation of the TLT in the Einstein-Maxwell-massless-scalar field theory, 
if such violations exist, is restricted to the narrow interval ${{eQ}\over{\hbar}}\in\big({1\over3},{1\over 2}\big)$. 

According to current knowledge, the lower bound of this range, $1/3$, can in principle be increased. 
This possibility, together with the arguments presented in the previous section, motivates us to conjecture that, 
within the classical framework of the coupled Einstein-Maxwell-massless-scalar field theory, extremal (zero-temperature) charged Reissner-Nordstr\"om black holes cannot form dynamically in the dimensionless regime 
\begin{equation}\label{Eq10}
{{eQ}\over{\hbar}}<{1\over2}
\qquad\Longrightarrow\qquad\text{No dynamically formed extremal black holes exist}. \
\end{equation}
Put differently, we conjecture that the classical Einstein-Maxwell-scalar field 
equations preserve the validity of the TLT in the regime where the vacuum remains quantum mechanically stable and quantum effects are unable to protect this fundamental principle. 

Our conjectured bound (\ref{Eq10}), which is physically motivated by the TLT, is in the spirit of, 
but slightly stronger than, the Schneider bound (\ref{Eq8}). 
In particular, the quantum validity of the TLT suggests that 
\begin{equation}\label{Eq11}
\Big({{eQ}\over{\hbar}}\Big)_{\text{min}}\geq{1\over2}\ .
\end{equation}
It would be interesting to know whether the exact bound is $({{eQ}/{\hbar}})_{\text{min}}=1/2$, 
precisely at the threshold beyond which quantum effects can no longer protect the TLT.

\bigskip
\noindent
{\bf ACKNOWLEDGMENTS}
\bigskip

We would like to thank Don Page for helpful correspondence. 



\begin{thebibliography}{99}
\bibitem{Noter0} Note that, in the small mass-radius regime $mR\ll\hbar$, most of the wave function 
is concentrated in the weak-gravity region $r\gg R$, well outside the central charged matter distribution. 
Hence, as discussed in \cite{Flam}, the spatial distribution of the electric charge within the effective radius $R$ contributes only at sub-leading order, namely $O[\ln^{-2}(\hbar/mR)]$, to the critical electric 
charge $({{eQ}/{\hbar}})_{\text{crit}}$ [see Eq. (\ref{Eq6})].
\end{thebibliography}
\end{document}